\documentclass[aps,twocolumn,superscriptaddress,prb]{revtex4}
\usepackage{amsmath}
\usepackage{color}
\usepackage{bbm}
\usepackage{amssymb}
\usepackage{epsfig}
\usepackage{multirow}
\usepackage{amsbsy}
\usepackage{array}
\usepackage{diagbox}
\usepackage{bm}
\usepackage{extarrows}
\usepackage{graphicx}
\usepackage{appendix}
\usepackage{txfonts}
\usepackage[normalem]{ulem}
\usepackage[colorlinks=true,linkcolor=blue,citecolor=blue,urlcolor=blue,bookmarks=false]{hyperref}

\begin{document}
	%\title{Nematic phase in periodically driven classical spins}
	%\title{Driving a two-dimensional frustrated Ising magnet with dipolar interactions:\\
	%	prethermal nematic order and staircase heating}
	%\title{Prethermal nematic order in a driven frustrated Ising magnet with dipolar interactions}
	\title{Prethermal nematic order and staircase heating in a driven \\frustrated Ising magnet with dipolar interactions}
	\date{\today}
	\author{Hui-Ke Jin}
	\affiliation{Department of Physics, Technische Universit{\"a}t M{\"u}nchen TQM, James-Franck-Stra{\ss}e 1, 85748 Garching, Germany}
	\author{Andrea Pizzi}
	\affiliation{Cavendish Laboratory, University of Cambridge, Cambridge CB3 0HE, United Kingdom}
	\affiliation{Department of Physics, Harvard University, Cambridge, Massachusetts 02138 USA}
	\author{Johannes Knolle}
	\affiliation{Department of Physics, Technische Universit{\"a}t M{\"u}nchen TQM, James-Franck-Stra{\ss}e 1, 85748 Garching, Germany}
	\affiliation{Munich Center for Quantum Science and Technology (MCQST), 80799 Munich, Germany}
	\affiliation{Blackett Laboratory, Imperial College London, London SW7 2AZ, United Kingdom}
	
	\begin{abstract}
		Many-body systems subject to a high-frequency drive can show intriguing thermalization behavior. Prior to heating to a featureless infinite-temperature state, these systems can spend an exponentially long time in prethermal phases characterized by various kinds of order. Here, we uncover the rich non-equilibrium phase diagram of a driven frustrated two-dimensional Ising magnet with competing short-range ferromagnetic and long-range dipolar interactions. We show that the ordered stripe and nematic phases, which appear in equilibrium as a function of temperature, underpin subsequent prethermal phases in a new multi-step heating process en route towards the ultimate heat death. We discuss implications for experiments on ferromagnetic thin films and other driving induced phenomena in frustrated magnets.
	\end{abstract}
	\maketitle

	\begin{figure*}[th]
		\includegraphics[width=\linewidth]{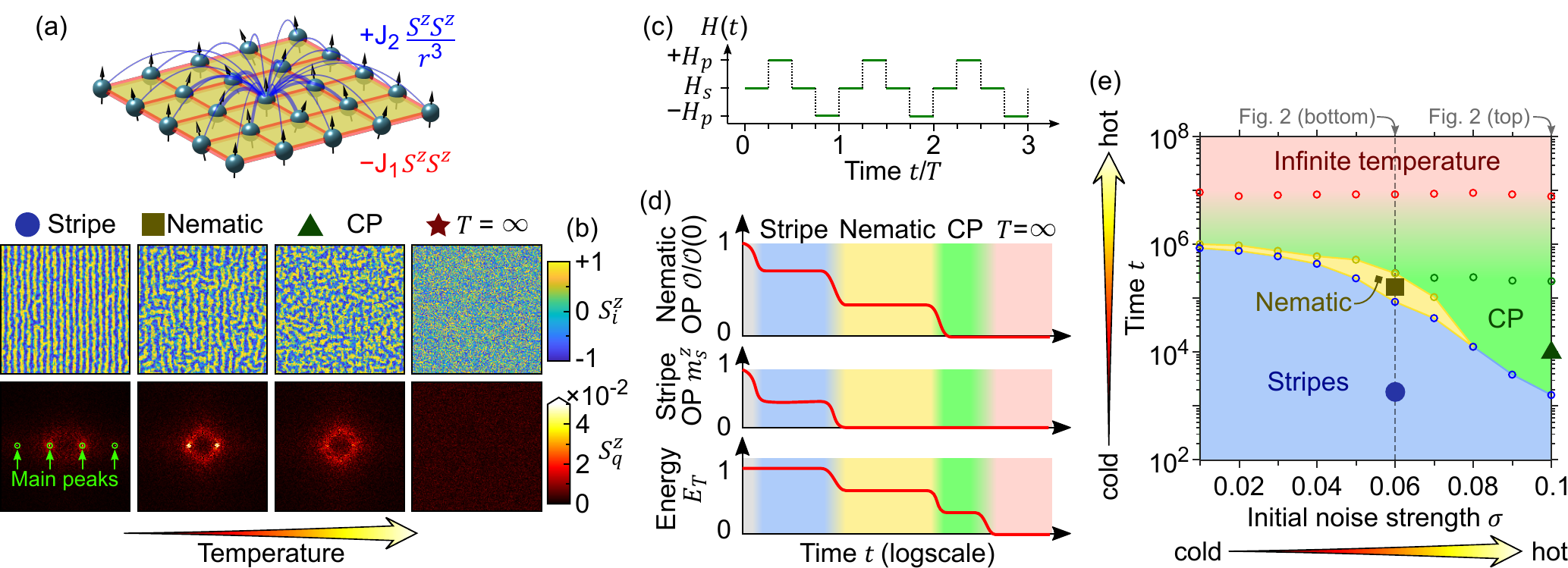}
		\caption{\textbf{A two-dimensional, driven, frustrated magnet}.
			(a) Spins on a two-dimensional lattice interact via ferromagnetic nearest-neighbour (red) and long-range antiferromagnetic (blue) Ising couplings.
			(b) As a function of temperature $T$, the model has several equilibrium phases shown with representative spin configurations both in real (top) and momentum (bottom) space. In order of increasing  $T$, we show the stripe, nematic, and correlated paramagnetic (CP) phases. Eventually, in the infinite temperature state the spins orientate randomly and become completely uncorrelated. 
			(c) The system is periodically driven by alternating the interactions in $H_s$ and a transverse field pulse in $H_p$, see Eq.~\eqref{eq:Ht}.
			(d) Schematic dynamics of the orientational and stripe order parameters $\mathcal{O}$ and $m_s^z$, respectively, and energy $E_T$. At high drive frequency $\omega$, the system can exhibit staircase heating: after an initial quick relaxation, the system first remains in a prethermal stripe regime ($m_s^z > 0$), then in a prethermal nematic one ($m_s^z = 0$, $\mathcal{O}>0$), and eventually in a CP ($m_s^z = \mathcal{O}= 0$, $E_T > 0$) that adiabatically heats up to infinite temperature ($E_T = 0$). The energy difference between stripe and nematic phases have been schematically amplified to stress this transition.
			(e) Non-equilibrium phase diagram in the plane of initial noise strength $\sigma$ and time $t$. The effective temperature of the system increases in the direction of both increasing $\sigma$, corresponding to higher initial effective temperatures, and $t$, corresponding to the dynamical heat up of the system through energy absorption. The dot, square, and triangle markers correspond to the configurations in (b). Here we used $L=72$ [$L=160$ in (b)], $J_1=1.55$ and $J_2=1$ ($W=4$), $g=0.46$, and $\omega=1.1$.
		}\label{fig:fig 1}
	\end{figure*}

	%{\em Introduction.}
	\section{Introduction}
	Frustrated magnetism has been a central research area in condensed matter physics for many decades~\cite{Vannimenus1977,Toulouse1987,BookDiep,BookLacroix}. Its defining feature is the presence of competing interactions leading to a (near) degeneracy of an extensive number of states. As a result, frustrated quantum models may host elusive quantum spin liquid ground states states with topological order~\cite{Lee08,Balents10,QSLRMP,Broholm2020,knolle2019field,savary2016quantum}. Their equilibrium finite-temperature behavior may also display remarkably rich physics stemming from the competition between small energy scales and entropy, for example crossovers into classical spin liquids~\cite{Wannier1950,Stephenson70}, liquid-gas like thermal phase transitions in spin ices~\cite{spinice,castelnovo2008magnetic}, the entropic selection of order via the order-by-disorder mechanism~\cite{orderbyd1,orderbyd2}, and different forms of nematic order~\cite{chandra1990ising,chalker1992hidden,mulder2010spiral,xu2008ising,fang2008theory,fernandes2014drives}. 
	
	More recently, non-equilibrium many-body physics has been a research area of immense interest partly because of the possibility to realise novel phases of matter beyond thermal equilibrium. As a prominent example, prethermal discrete time crystals (DTCs) have been recently predicted~\cite{Else2017,Machado2020,pizzi2021higher} and observed~\cite{Kyprianidis2021} to break the time translational symmetry of a periodic Floquet drive over very long transient times. At the core of prethermal DTCs is the phenomenon of prethermalization~\cite{Berges2004,Bukov2015, Mori2016,Canovi2016,Weidinger2017,Abanin2017CMP,Abanin2017,Mallayya2019,Luitz2020,Zhao2021}, whereby energy absorption from a periodic drive is suppressed for large enough drive frequencies $\omega$. These systems equilibrate to an effective thermal state with respect to an effective Hamiltonian and an effective temperature set by the initial states. From the viewpoint of driven magnetism, a prethermal DTC can be understood as a simple ordered ferromagnetic phase (of an unfrustrated effective Hamiltonian) at stroboscopic times. After an exponentially long time $\sim e^{c\omega}$ with $c$ a constant, these systems melt into a featureless state akin to infinite temperature. Requiring only a high frequency drive, prethermalization is ideal for the investigation of non-equilibrium many-body physics in experiments ~\cite{Gring2012,Langen2013,Beatrez2021}.
	
	Initially studied within a quantum formalism, prethermalization and prethermal DTCs have recently been established to also emerge classically~\cite{Rajak2018,Mori2018,Rajak2019,Howell2019,Andrea2021,Andrea2021PRB,Ye2021}, which greatly simplifies the investigation of these phenomena both beyond one-dimensional examples as well as in the presence of long-range interactions. With the possibility to explore the interplay of interaction range and dimensionality at hand, a natural question is whether novel prethermal phases as well as non-trivial heating dynamics may exist in many-body systems with competing interactions.
	
	Here, we investigate prethermalization in a driven frustrated magnet. Concretely, we study a two-dimensional Ising model with competing ferromagnetic short-range and dipolar-like long-range interactions, see Fig.~\ref{fig:fig 1}(a). The static model was first put forward as a minimal description of ultrathin magnetic films~\cite{Stampanoni1987,Pescia1987,Krebs1988,Allenspach1990,Pappas1990,Allenspach1992,Qiu1993,Vaterlaus2000}, in which the interplay of frustration and thermal fluctuations gives rise to a rich phase diagram hosting various kinds of magnetic orders~\cite{Booth1995,Abanov1995,MacIsaac1995,DeBell2000,Canna2004,Cannas2005,Cannas2006,Lucas2007,Pighin2007}. As shown in Fig.~\ref{fig:fig 1}(b), it includes (in order of increasing temperature) a magnetic stripe phase with long-ranged orientational and magnetic order, a nematic phase that only breaks the lattice rotational symmetry, and a correlated paramagnetic (CP) regime preserving all of the allowed symmetries while displaying characteristic short-range correlations. More recently, magnetic thin films have also emerged as ideal experimental platforms for investigating non-equilibrium  magnetic phenomena~\cite{Barman2007,Pfau2012,Lambert2014,Yu2016,Yu2017,Hrabec2017,trager2021real}, for example realizing transient topological defects after ultrafast laser excitation~\cite{Buttner2021}.

	\begin{figure*}[!ht]
		\includegraphics[width=1\linewidth]{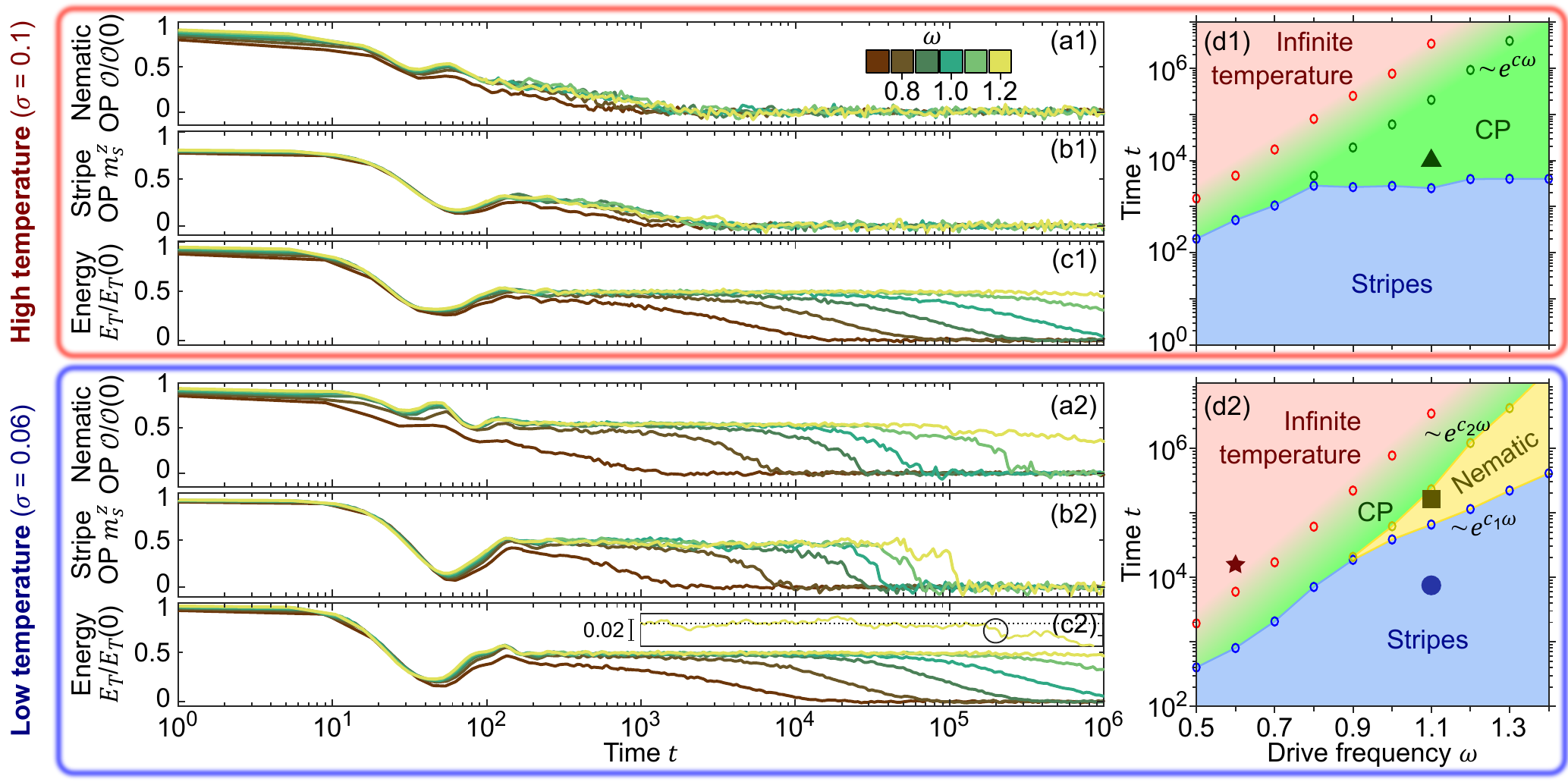}
		\caption{{\bf Diagnosing prethermal order.} From the temporal profiles of the order parameters at various drive frequencies $\omega$ (left) we map out the system's non-equilibrium phase diagrams (right). For a high-temperature initial condition (top, $\sigma=0.1$) the orientational order parameter $\mathcal{O}$ (a1) and stripe magnetization $m^z_s$ (b1) quickly and simultaneously decay to zero, while prethermalization is still imprinted on the relative energy $E_T$ (c1), reaching its infinite-temperature value only after an exponentially long time $\sim{}e^{c\omega}$. The phase diagram (d1) shows an initial transient stripe regime followed a prethermal CP regime, beyond which the infinite-temperature regime is reached. In contrast, for low-temperature initial states (bottom, $\sigma = 0.06$), the system exhibits two prethermal plateaus: at large enough drive frequency $\omega$,  $\mathcal{O}$ (a2) and $m^z_s$ (b2) remain finite over two timescales $\sim{}e^{c_2\omega}$ and $\sim{}e^{c_1\omega}$, respectively and with $c_2>c_1$, indicating subsequent stripe and nematic prethermal phases. The energy (c2) also decays to $0$ after an exponentially large time, and features two-plateau structure corresponding to the stripe to nematic transition emerging at large $\omega$, as highlighted by the circle in the inset. The stripe and nematic prethermal phases characterize the dynamical phase diagram (d2). Note that the blue regime in (d1) and green regime in (d2) persist in a transient rather than prethermal manner.
			In (d1,d2), the markers correspond to the parameters used in plotting the spin configurations in Fig.~\ref{fig:fig 1}(b). Here, $L=72$, $J_1=1.55$ and $J_2=1$ ($W=4$), and $g=0.46$.}\label{fig:fig 2}
	\end{figure*}
	
	Turning to the driven setting for such a frustrated magnet, we uncover a rich non-equilibrium phase diagram.  For initial states corresponding to low effective temperatures, we find a remarkable staircase heating process induced by the competition between subsequent prethermal stripe and nematic phases. In the regime of high initial effective temperature, this multi-step heating reduces to a one-step process and the prethermal nematic phase transforms to a symmetric CP phase. Finally, we discuss how these phenomena could be experimentally observed in magnetic thin films and speculate about other non-equilibrium phenomena in driven frustrated magnets.

	The rest of this paper is organized as follows. In Sec.~\ref{sec:model}, we present the dynamical Hamiltonian, formulate the equations of motion, and introduce observables and initial states. We discuss our main results in Sec.~\ref{sec:rec}. Finally, Sec.~\ref{sec:summary} is devoted to a brief summary and discussion.
	
	%{\em Model.}
	\section{Model}\label{sec:model}
	We consider a system of $L^2$ classical spins $\bm{S}_i = \left(S_i^x,S_i^y,S_i^z\right)$, arranged on a square lattice of linear size $L$ and described by the following static Hamiltonian
	\begin{align}
		H_s=-\frac{J_1}{4}\sum_{\langle{}ij\rangle}{S}^{z}_i{}{S}^z_{j}+\frac{J_2}{\mathcal{N}}\sum_{i\neq{}j}\frac{1}{r_{i,j}^{3}}{S}^{z}_i{}{S}^z_{j},\label{eq:staticH}
	\end{align}
	where $J_1$ and $J_2$ are the nearest-neighbor ferromagnetic and dipolar couplings, respectively, and $\mathcal{N}=\sum_{i=1}^{N}(r_{1,i})^{-3}$ is a Kac-like normalization factor. The $O(3)$ spins $\bm{S}_i$ with unitary modulus can be parameterized by a pair of polar and azimuthal angles $(\theta_i,\phi_i)$ as $\bm{S}_i=\left(\sin\theta_i\cos\phi_i,\sin\theta_i\sin\phi_i,\cos\theta_i\right)$. The distance $r_{i,j}$ between sites $i$ and $j$ is taken to be $r_{ij}=\sum_{a=x,y}\frac{L_a}{\pi}\left|\tan\left(\frac{\pi(i_a-j_a)}{L_a}\right)\right|$, which implicitly implements periodic boundary conditions and reduces finite-size effect. Eq.~\eqref{eq:staticH} has two symmetries: the $Z_2$ symmetry associated to spin flips $S_j^z \to -S_j^z$, and the $C_4$ symmetry associated to lattice rotations by $90$ degrees.
	
	The presence of these symmetries gives rise to a rich equilibrium phase diagram. At temperatures much larger than the bare exchange constants, the system develops characteristic short-range correlations while preserving all symmetries in the CP crossover regime. At zero temperature, the energy in Eq.~\eqref{eq:staticH} is minimized for spins aligned along $z$ axis, which connects our model to the known results for Ising variables~\cite{Booth1995,Abanov1995,MacIsaac1995,Cannas2005,Cannas2006,Lucas2007}.
	In the absence of the short-range interaction, $J_1=0$, the system is in an antiferromagnetic phase that breaks the $Z_2$ symmetry while fulfilling the $C_4$ one. When $J_1$ and $J_2$ are comparable, the spins arrange in stripes that, on top of the $Z_2$ symmetry, also break the $C_4$ lattice rotational symmetry down to a $C_2$ one~\cite{MacIsaac1995}. The width $W$ of the stripes scales as $e^{J_1/J_2}$, and the transition between the stripe and antiferromagnetic phases is found, for $T \to 0$, at $J_1/J_2 \approx 0.2$. Remarkably, at a moderately large temperature a third phase emerges, the nematic one, where the stripe magnetization is lost but nematic (meaning orientational) order persists~\cite{Cannas2005,Lucas2007}.
	
	We are interested in the study of the interplay between competing short- and long-range interactions in the non-equilibrium realm. To this end, we consider a periodically driven version of the model in Eq.~\eqref{eq:staticH}, where the Hamiltonian $H_s$ is alternated with transverse field pulses according to the following quaternary Floquet Hamiltonian, at frequency $\omega=2\pi/\tau$ [see Fig.~\ref{fig:fig 1}(c)]
	\begin{equation}
		H(t) =
		\begin{cases}
			H_s & {\rm~for~} t\in[0,~\tau/4)\\
			H_p = +2\omega{}g\sum_{i}{S}^{x}_i{} & {\rm~for~} t\in[\tau/4,~\tau/2)\\
			H_s & {\rm~for~} t\in[\tau/2,~3\tau/4)\\
			-H_p & {\rm~for~} t\in[3\tau/4,~\tau).
			%-2\omega{}g\sum_{i}{S}^{x}_i{} & {\rm~for~} t\in[\frac{3T}{4},T),
		\end{cases}
		\label{eq:Ht}
	\end{equation}
	The effect of the transverse field is to rotate the spins around $x$ axis by an angle $\pi g$ (twice in each period, in opposite directions) which, given the parametrization $\omega g$ of the field strength, is irrespective of the frequency $\omega$. The spin dynamics is described by standard Hamilton equations $\partial_t \bm{S}_i=\{\bm{S},H(t)\}$, and can be analytically integrated as~\cite{Howell2019,Andrea2021}
	\begin{align*}
		\bm{S}_i(n\tau+\tau)=\left(\begin{array}{ccc}
			c_1 & -s_1 & 0\\
			s_1c_2 & c_1c_2 & s_2\\
			-s_1s_2 & -c_1s_2 &c_2\\
		\end{array}\right)
		\left(\begin{array}{ccc}
			c_1 & -s_1 & 0\\
			s_1c_2 & c_1c_2 & -s_2\\
			s_1s_2 & c_1s_2 &c_2\\
		\end{array}\right)
		\bm{S}_i(n\tau),
	\end{align*}
	where $c_1=\cos(\kappa_{i}\tau/4)$, $s_1=\sin(k_{i}\tau/4)$, $c_2=\cos(\pi{}g)$, and $s_2=\sin(\pi{}g)$. 
	The effective field $\kappa_i$ reads $\kappa_i=-\frac{J_1}{4}\sum_{j\in\partial_i}S^z_j+\frac{J_2}{\mathcal{N}}\sum_{j\neq{}i}\frac{1}{r_{ij}^3}S^z_j,$ where $\sum_{j\in\partial_i}$ denotes summation over the nearest neighbors of site $i$. We note that our choice of a step-wise driving protocol is motivated by computational convenience for probing long times but the following results are expected to be similar for more general choices of continuous drives~\cite{pizzi2021higher}. 
	
	We initialize the system in a stripe state with $Y$-orientation. For concreteness, we henceforth set $J_1=1.55$ and $J_2=1$, where the ground state is indeed a stripe state with width $W=4$ (see Appendix~\ref{sec:eqphase}). The spins are initialized in such a ground state. On top of this, we add a perturbation that brings the many-body character of the system into play. Specifically, $\theta_i$'s are perturbed using a Gaussian noise, $p(\theta)=\frac{1}{\sqrt{2\pi}\tilde{\sigma}}e^{-\theta^2/2\tilde{\sigma}^2}$ with $\tilde{\sigma}\equiv{}2\pi\sigma$ the standard deviation, whereas $\phi_i$'s are randomly drawn in $[0,2\pi]$. The parameter $\sigma$ thus acts as a sort of initial temperature, injecting excitations on top of the ground state. Note that this choice of initial states is convenient for simulations but in an experimental realization a low temperature stripe state would be sufficient. 
	
	The different possible dynamical regimes can be diagnosed with suitable observables. Generalizing the idea proposed for Ising variables in Ref.~\cite{Booth1995}, we consider the following orientational order for $O(3)$ spins
	\begin{align}
		\mathcal{O}(t)=\frac{\sum_{i}\bm{S}_{i}(t)\cdot\bm{S}_{i+\hat{y}}(t)-\bm{S}_{i}(t)\cdot\bm{S}_{i+\hat{x}}(t)}{2N-\sum_{i}{}\bm{S}_{i}(t)\cdot\bm{S}_{i+\hat{x}}(t)-\bm{S}_{i}(t)\cdot\bm{S}_{i+\hat{y}}(t)}.
	\end{align}
	The parameter $\mathcal{O}$ is equal to $+1$ $(-1)$ in a perfect stripe state with $Y$ ($X$) orientation, and vanishes if the $C_4$ symmetry is preserved. Note that $\mathcal{O}(t)$ results rather sensitive to perturbations when the alignment of $\bm{S}_{i}$ and $\bm{S}_{j}$ is slightly spoiled. For $\sigma=0.06$, the initial condition gives $\mathcal{O}(0)\approx0.5$, while still exhibiting a clear stripe structure. Therefore, we find $\mathcal{O}(t)$ to be more informative when normalized as $\mathcal{O}(t)/\mathcal{O}(0)$. To further characterize the magnetic order, we then consider the stripe magnetization $\bm{m}_s(t)=\frac{1}{N}\sum_{i}(-1)^{\left(i_x/W~\mbox{mod}~2\right)}\bm{S}_i(t)$,
	which detects stripe spin configurations with width $W$. Finally, to keep track of energy absorption in the system, we look at the normalized average energy per period, $E_{T}(t)=H_s(t)/H_s(t=0)$  (with $H_s(t=0)\approx{}-0.61L^2$ for $\sigma=0.06$ and $L\geq{}72$). All these observables are computed at stroboscopic times $t=\tau,2\tau,3\tau,\dots$. Note, we also consider higher-order expansions for the prethermal effective Hamiltonian in Appendix~\ref{sec:highfexp}.
	
	%{\em Results.}
	\section{Results}\label{sec:rec}
	According to the prethermalization paradigm, at large drive frequency $\omega$ the system can remain stuck in a prethermal regime for an exponentially long time $\sim e^{c\omega}$. The dynamical response of the system crucially depends on the amount of excitations injected in the initial condition, that is, on the system's initial effective temperature. Specifically, we distinguish the regimes of low and high initial effective temperatures, that is small and large $\sigma$, analysed in the top and bottom panels of Fig.~\ref{fig:fig 2}, respectively.
	
	For a high initial effective temperature (e.g., $\sigma = 0.1$), the initial stripe configuration is unstable, and quickly melts into a CP phase: both the orientational and stripe order parameters quickly decay to $0$, see Fig.~\ref{fig:fig 2}(a1,b1). Such a transition occurs over a timescale $\sim 1/\lambda$, with $\lambda$ the Lyapunov exponent associated to the effective Hamiltonian $H_s$, which becomes independent of $\omega$ in the high frequency limit~\cite{Andrea2021}. On the other hand, the high drive frequency hinders energy absorption, see Fig.~\ref{fig:fig 2}(c1). Indeed, even in the absence of symmetry breaking in the prethermal regime, the infinite-temperature state with $E_T = 0$ is only reached after an exponentially long time $\sim e^{c\omega}$.
	
	The situation changes substantially for an initial condition with low effective temperature (e.g., $\sigma = 0.06$). The system not only breaks a symmetry in the prethermal regime, but it does so twice: first the system prethermalizes to a stripe state for a time $\sim e^{c_1 \omega}$, and then to a nematic phase for a time $\sim e^{c_2\omega}$, before reaching the infinite-temperature state, see Fig.~\ref{fig:fig 2}(a2-c2). (i) A first prethermal stripe phase, extending over a timescale $\sim{}e^{c_1\omega}$, is signaled by the equilibration of $m^{y}_s$, $m^{z}_s$, $E_T$, and $\mathcal{O}$ to a finite prethermal value, that does not depend on the drive frequency $\omega$ (note, $m^{x}_s \approx 0$). (ii) At later times, and provided the frequency is large enough ($\omega \gtrapprox 0.9$), the stripe order parameters $m^{y}_s$ and $m^{z}_s$ decay to $0$, while the energy $E_T$ and the nematic order parameter $\mathcal{O}$ drop slightly while remaining finite. This signals the second prethermal phase, the nematic one, extending for a time $\sim e^{c_2 \omega}$, with $c_2>c_1$. From our data, we estimate the prethermal energy plateau for the stripe and nematic phases at $E_T/E_T(0) = 0.4894\pm0.0082$ and $0.4672\pm0.0106$, respectively. The small extent of this drop, which we try to highlight with a circle in the inset of Fig.~\ref{fig:fig 2}(c2), is related to the narrow temperature window of the nematic phase in the equilibrium phase diagram of the static Hamiltonian in Eq.~\eqref{eq:staticH}. Indeed, this non-equilibrium evolution can be understood from the equilibrium phase diagram of $H_s$ by doing the association ``time'' $\leftrightarrow$ ``temperature'' (see the equilibrium phase diagram in Appendix~\ref{sec:eqphase}). (iii)  After undergoing a CP crossover, the system eventually reaches the infinite-temperature state, signaled by the observables of interest reaching their infinite-temperature value $0$. 
	
	From the time traces of the observables described above, we map out the dynamical phase diagrams of the system in Fig.~\ref{fig:fig 2}(d1,d2), drawn in the plane of drive frequency $\omega$ and time $t$. Specifically, we consider that the system leaves the stripe phase when $m_s^z$ crosses the value $10^{-2}$ (blue circles) and the nematic phase when $\mathcal{O}/\mathcal{O}(0)$ crosses $10^{-2}$ (yellow circles). Furthermore, for the CP phase, we use green and red circles to indicate when $E_T$ crosses the thresholds $0.45$ and $0.005$, respectively.
	
	A further quantity convenient for diagnosing the various phases is the structure factor $S^a_{\bm{q}} = \left|\frac{1}{N}\sum_jS^a_i{}e^{i\bm{q}\cdot{}\bm{j}}\right|$ ($a=x,y,z$).
	Fig.~\ref{fig:fig 1}(d) illustrates several typical spin configurations in $z$ axis and corresponding $S^{z}_{\bm{q}}$ for different phases. Note that the spin configurations in $x$ axis are always disordered during the drive process, whereas those in $y$ axis are very similar to the one in $z$ axis. 
	In the $Y$-oriented stripe phase with $W=4$, $S^z_{\bm{q}}$ exhibits four sharp peaks at $\bm{q}=(\frac{2w+1}{4},0)\pi$ with $w=0,1,2,3$. In the nematic phase, only two broader peaks remain, signalling breaking of the $C_4$ symmetry, while a light ring-shaped feature around the origin emerges. For a typical CP state preserving the lattice rotational symmetry, all of the sharp peaks disappear and only the ring-shaped feature is left. 
	
	In Fig.~\ref{fig:fig 1}(e), we show the non-equilibrium phase diagram as a function of initial noise strength $\sigma$, at a fixed drive frequency $\omega = 1.1$. When $\sigma$ is very small, the stripe phase is essentially the only prethermal phase. For moderate $\sigma$ (e.g., $\sigma\approx{}0.06$), a long-lived prethermal nematic phase emerges and gives rise to clear staircase heating with timescales $\sim{}e^{c_1\omega}$ and $\sim{}e^{c_2\omega}$. At even larger $\sigma\gtrapprox0.08$, stripe order persists just in a transient (rather than prethermal) fashion, whereas nematic order is not observed. Instead, a prethermal CP phase emerges. The non-equilibrium phase diagram can be understood in analogy with the equilibrium one by noting that the effective temperature of the system can be increased either in the form of larger initial noise $\sigma$, or by just waiting longer through energy absorption. Note, we have also verified the stability of the prethermal phases to generic small perturbations of the driving protocol, e.g., to the addition of random $Z$-fields to $H_s$ in Eq.~\eqref{eq:staticH}.
	
	%{\em Discussion and conclusion.}
	\section{Discussion and conclusion}\label{sec:summary}
	Efficiently integrated over long timescales and for large system sizes, the microscopic equations of motion for classical systems are a new tool to study prethermalization in higher dimension.
	For instance, focusing on a two-dimensional Ising magnet with competing short- and long-range interactions, we find an unprecedented two-step prethermalization process, consisting of subsequent prethermal stripe and nematic phases for times $\sim e^{c_1 \omega}$ and $\sim e^{c_2 \omega}$, respectively, before the ultimate heath death. 
	A difference in the two scaling coefficients, $c_1\neq{}c_2$, should be attributed to the fact that the respective phase transitions are driven by the proliferation of different kinds of defects, i.e., the dislocations of stripes melt $m^z_s$ and the proliferations of domain walls between different stripe orientations eventually disorder the nematic state.
	
	We note that the physical picture that we outlined is expected to persist in the presence of small symmetry-breaking terms added to $H_s$, and taken with different signs in the first and third quarters of the driving protocol in Eq.~\eqref{eq:Ht}. In this case, the stripe and nematic prethermal phases would break a symmetry not of $H_s$, which in fact would have none, but of an effective prethermal Hamiltonian instead, thus signalling the genuine non-equilibrium nature of these phenomena similar to the case of prethermal DTCs~\cite{Kyprianidis2021,Andrea2021,Ye2021}.
	
	An intriguing prospect from our proposal is the experimental realization in magnetic thin film materials. Our model Hamiltonian is an effective description of a whole class of perpendicular magnetic anisotropy (PMA) materials~\cite{Dieny2017}. In particular, to avoid rapid heating via coupling to charge excitations, the recently discovered insulating PMA materials appear to be the most promising~\cite{Kocsis2016,Mogi2018,Csizi2020,Laith2021}. These can develop stripe domains from the interplay between a relatively large easy-axis anisotropy and dipole-dipole interactions. The relevant excitation energy scale is that associated to the anisotropy, $E_{A}\sim{}1$ meV~\cite{Yang2011}, which compares favourably with the available sub Thz drives meeting the condition $\omega{}>{}E_A$ necessary for prethermalization. Moreover, discerning the various prethermal phases should be facilitated by the recent advances in the real-time imaging of magnetic domains in PMA systems~\cite{Miguel2006,Kronseder2015,Voltan2016}.
	
	In the future, it would be interesting to investigate how the prethermal phase diagram outlined here changes for different initial conditions beyond the considered noisy stripe states. More broadly, a worthwhile direction for future research is the study of prethermal order-by-disorder mechanisms in frustrated magnets, with the possibility to stabilize selected magnetic phases by different choices of the drive. Another natural open question is whether driven frustrated magnets of the kind studied here could lead to new forms of time crystalline order. In general, we expect driven frustrated magnets to be a versatile playground for novel non-equilibrium phenomena. 
	
	\begin{acknowledgments}
		We thank Michael Knap for helpful feedback on the manuscript. We thank Istvan Kezsmarki and Christian Back for illuminating discussions about the experimental feasibility of prethermal phenomena in PMA materials.  
		H.-K.~J.~is funded by the European Research Council (ERC) under the European Unions Horizon 2020 research and innovation program (Grant agreement No.~771537). A.~P.~acknowledges support from the Royal Society and hospitality at TUM. J.~K.~acknowledges support from the Imperial- TUM flagship partnership. The research is part of the Munich Quantum Valley, which is supported by the Bavarian state government with funds from the Hightech Agenda Bayern Plus.
	\end{acknowledgments}
	
	\appendix
	
	\section{Equilibrium phase diagram}\label{sec:eqphase}
	
	\begin{figure}[htp]
		\centering
		\includegraphics[width=1.0\linewidth]{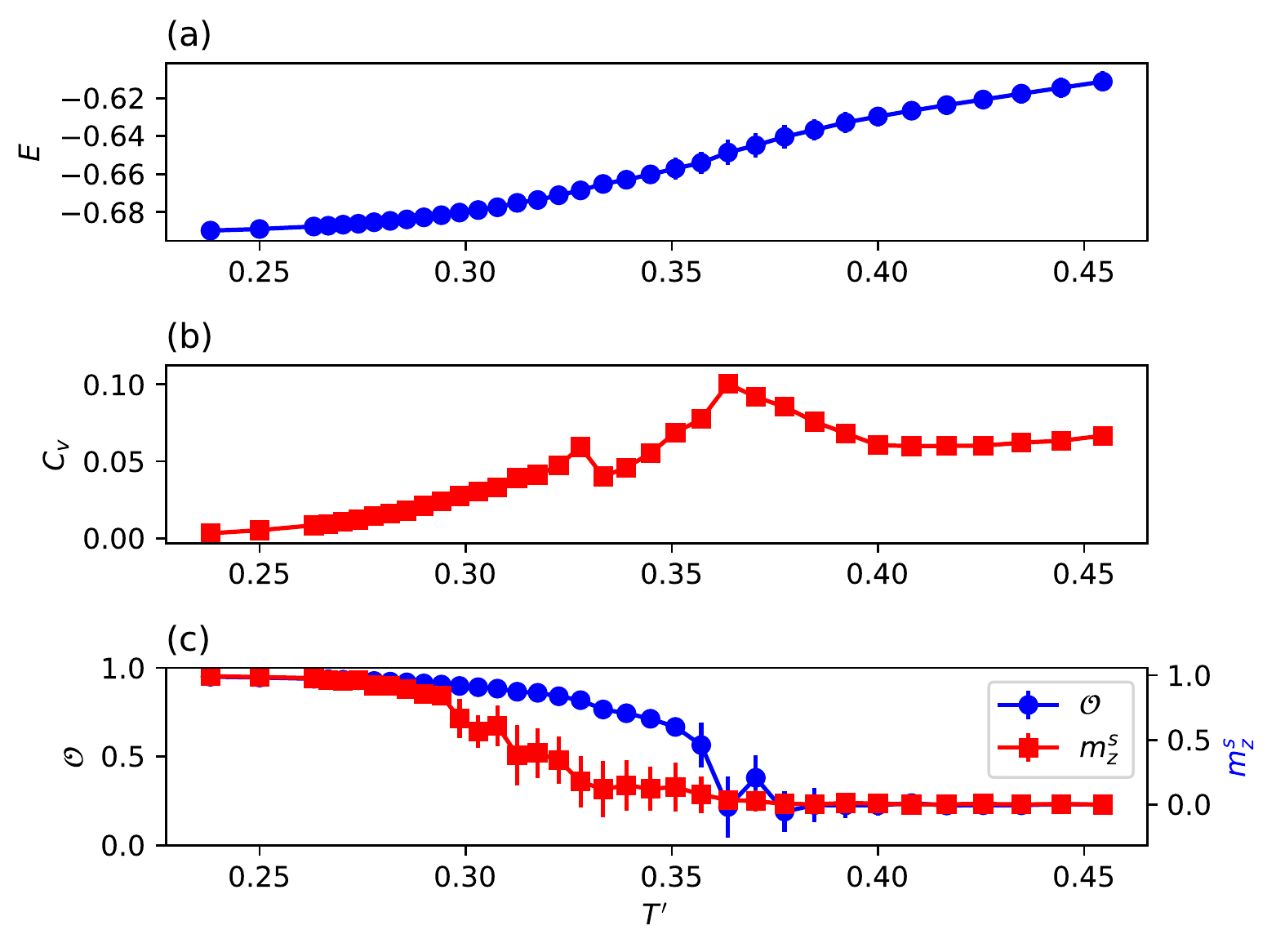}
		\caption{\textbf{Equililbrium phase diagrm of $H_s$.} We plot the free energy density (a), specific heat (b), and stripe as well as nematic orders (c) at equilibrium and versus the temperature $T^\prime$, as obtained by Monte Carlo simulations. Here, we consider $L = 72$ and $2\times{}10^7$ Monte Carlo steps to measure the observables. To reduce the autocorrelation of the samples, we have used only one sample for measurements every ${72}^2$ Monte Carlo steps.}
		\label{fig:MCdata}
	\end{figure}
	
	\begin{figure*}[thp]
		\centering
		\includegraphics[width=1\linewidth]{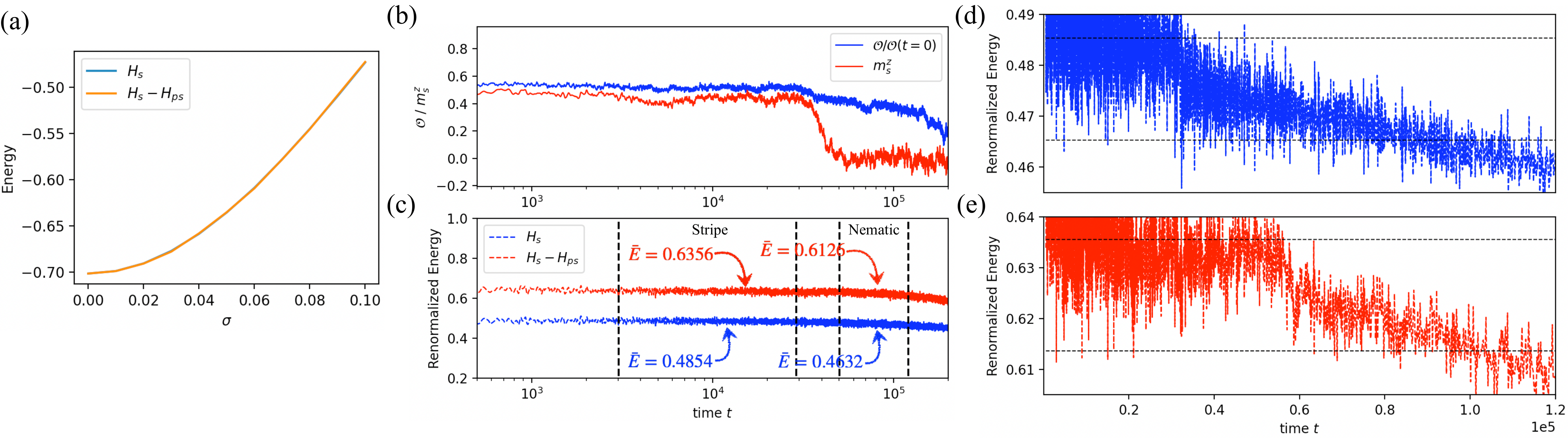}
		\caption{
			(a) Energy of the initial stripe state as a function of Gaussian noise strength $\sigma$ for prethermal Hamiltonians $\frac{1}{2}H_s$ (blue) and $\frac{1}{2}H_s-\frac{1}{2}H_{ps}$ (orange). Here, the lattice size is L=72 and the results are ensemble averaged over 100 different realizations.
			(b) Dynamics of the stripe order $m^z_s$ and nematic order $\mathcal{O}$. (c) Dynamics of the zero-th prethermal Hamiltonian $H_s$ and higher-oder prethermal Hamiltonian $H_s-H_{ps}$. The mean value of each energy plateau is obtained as an average over the time window indicated by black dashed lines. To better emphasize the energy difference between stripe and nematic phases, we plot in (d)-(e) their zoomed version.
			Here, $L=72$, $J_1=1.55$ and $J_2=1$ ($W=4$), $\omega=1.1$, and $g=0.46$.}
		\label{fig:engs_higherorder}
	\end{figure*}
	
	We implement classical Monte Carlo simulations to obtain the equilibrium phase diagram for the static Hamiltonian $H_s$ at the point of interest $J_1=1.55$ and $J_2=1$. 
	Specifically, we investigate the free energy $E=\langle{}H_s\rangle$, heat capacity $C_V=\langle{}H_s^2{}\rangle-\langle{}H_s\rangle{}^2$, stripe order parameter $m_s^z$, and nematic order parameter $\mathcal{O}$ as a function of the temperature $T^\prime$, in Fig.~\ref{fig:MCdata}. (i) At low temperature $T^\prime \lessapprox 0.33$, we observe a stripe phase ($m_s^z\approx{}1$ and $\mathcal{O}\approx{}1$). (ii) When $0.33 \lessapprox T^\prime \lessapprox 0.36$, the system enters a nematic phase ($m_s^z\approx{}0$ and $\mathcal{O}>0$). (iii) When $T^\prime \gtrapprox 0.36$, we find a correlated paramagnetic (CP) phase ($m_s^z=\mathcal{O}=0$), which adiabatically connects to infinite temperature $T^\prime\rightarrow\infty$. The peaks of the speciﬁc heat $C_V$ signal two second order phase transitions. A few observations are in order. (i) The stripe-nematic transition is less evident than the nematic-paramagnetic one; (ii) The temperature range corresponding to the nematic phase is very narrow, its free energy varying only from 0.66 to 0.64, which is consistent with the small energy difference between stripe and nematic phases in Fig.~\ref{fig:fig 2} in the main text; (iii) The CP phase does not break any symmetry, it is just a disordered phase that retains a finite free energy $E$ and, correspondingly, short range correlations --- this motivates the nomenclature ``CP phase'', consistently with previous literature.

	\section{High frequency expansion}\label{sec:highfexp}
	
	Below we consider a higher order expansion of prethermal Hamiltonian in the high-frequency limit. The equation of motion, $d{}\vec{S}_i/dt=\{H(t),\vec{S}_i\}$, leads to a stroboscopic evolution function,
	\begin{equation*}
		\vec{S}_i(N\tau)=e^{\frac{\tau}{4}\{-H_p,~\cdot\}}e^{\frac{\tau}{4}\{H_s,~\cdot\}}e^{\frac{\tau}{4}\{H_p,~\cdot\}}e^{\frac{\tau}{4}\{H_s,~\cdot\}}
		\vec{S}_i (N\tau-\tau).
	\end{equation*}
	We can define the effective prethermal Hamiltonian $H_{\rm eff}$ as 
	\begin{equation*}
		e^{\frac{\tau}{4}\{-H_p,~\cdot\}}e^{\frac{\tau}{4}\{H_s,~\cdot\}}e^{\frac{\tau}{4}\{H_p,~\cdot\}}e^{\frac{\tau}{4}\{H_s,~\cdot\}}=e^{\tau\{H_{\rm eff},~\cdot\}}.
	\end{equation*}
	In the high frequency limit, we can expand the above equation by using Baker-Campbell-Hausdorff (BCH) formula, e.g.,
	\begin{equation}
		e^{\frac{\tau}{4}\{H_p,~\cdot\}}e^{\frac{\tau}{4}\{H_s,~\cdot\}}\approx{}e^{\frac{\tau}{4}\{H_p+H_s+H_{ps},~\cdot\}},
	\end{equation}
	where the explicit form of $H_{ps}\equiv{}\frac{\tau}{8}\{H_p,H_s\}$ reads
	\begin{equation*}
		H_{ps}=\frac{\tau\omega{}g}{4}\left[\frac{J_1}{4} \sum_{\langle{}ij\rangle}(S_i^z S_j^y+S_i^y S_j^z)-\frac{J_2}{\mathcal{N}}\sum_{i\neq{}j}\left( \frac{1}{(r_{ij}^3}(S_i^z S_j^y+S_i^y S_j^z \right)\right].
	\end{equation*}
	We then obtain that
	\begin{equation}
		e^{\tau\{H_{\rm eff},~\cdot\}}\approx{}e^{\frac{\tau}{4}\{-H_p+H_s-H_{ps},~\cdot\}}e^{\frac{\tau}{4}\{H_p+H_s+H_{ps},~\cdot\}},    
	\end{equation}
	which leads to 
	\begin{equation}
		H_{\rm eff}\approx{}\frac{1}{2}H_s-\frac{1}{2}H_{ps}+\frac{\tau}{16}\{H_s, H_{ps}\}+O(\tau^2).
	\end{equation}
	By noting that $\tau{}{\omega}=2\pi$, $H_{ps}$ does not depend on drive frequency $\omega$ and $\tau{}\{H_s, H_ps\}$ indeed is linear in $1/\omega$. The emergence of $H_{ps}$ is consistent with the results we have observed in the prethermal phases, i.e., both the $y$ and $z$ components of spin exhibit ordered (striped and nematic) configurations, but the $x$ component is disordered.
	
	In Fig.~\ref{fig:engs_higherorder}(a), we evaluate the energy of our initial stripe state with the prethermal Hamiltonian $H_{\rm eff}=\frac{1}{2}H_s-\frac{1}{2}H_{ps}$, and find the contribution from the latter term, $\frac{1}{2}H_{ps}$, is negligible.
	We also study the evolutions of energy evaluated with Hamiltonians $H_s$ and $H_s-H_{ps}$, respectively, as show in Fig.~\ref{fig:engs_higherorder}(c). We find that, after entering the prethermal stripe and nematic phases, the higher-order corrections to the energy become more significant. Nonetheless, the relative energy difference for the higher-order prethermal Hamiltonian $H_s-H_{ps}$, $\approx{}3.6\%$, is still very small, and comparable with that for the zeroth-order prethermal Hamiltonian $H_s$, $\approx{}4.6\%$. We therefore argue that the energy difference is small is ultimately due to the narrowness of the temperature range of the nematic phase in the equilibrium phase diagram, irrespective of the order of the expansion of the prethermal Hamiltonian $H_s$.

	\bibliography{DN.bib}
	
\end{document}